\documentclass[aps,prd,groupedaddress,amssymb,twocolumn,eqsecnum,showpacs,epsfig,nofootinbib]{revtex4}

\usepackage{graphicx}
\usepackage{bm}
\usepackage{dcolumn}
\usepackage{amsmath}

\usepackage{amsmath}
\usepackage{amssymb}

\numberwithin{equation}{section}
\usepackage{epsfig}

\begin{document}
\newcommand{\newc}{\newcommand}

\newc{\be}{\begin{equation}}
\newc{\ee}{\end{equation}}
\newc{\ba}{\begin{eqnarray}}
\newc{\ea}{\end{eqnarray}}
\newc{\bea}{\begin{eqnarray*}}
\newc{\eea}{\end{eqnarray*}}
\newc{\D}{\partial}
\newc{\ie}{{\it i.e.} }
\newc{\eg}{{\it e.g.} }
\newc{\etc}{{\it etc.} }
{\newc{\etal}{{\it et al.}}
\newc{\lcdm}{$\Lambda$CDM}
\newcommand{\nn}{\nonumber}
\newc{\ra}{\rightarrow}
\newc{\lra}{\leftrightarrow}
\newc{\lsim}{\buildrel{<}\over{\sim}}
\newc{\gsim}{\buildrel{>}\over{\sim}}
\newcommand{\mincir}{\raise
-3.truept\hbox{\rlap{\hbox{$\sim$}}\raise4.truept\hbox{$<$}\ }}
\newcommand{\magcir}{\raise
-3.truept\hbox{\rlap{\hbox{$\sim$}}\raise4.truept\hbox{$>$}\ }}

\title{The generalized evolution of linear bias: a tool to test gravity}

\author{S. Basilakos}\email{svasil@academyofathens.gr}
\affiliation{Academy of Athens, Research Center for Astronomy and
Applied Mathematics,
 Soranou Efesiou 4, 11527, Athens, Greece}

\author{M. Plionis}\email{mplionis@astro.noa.gr}
\affiliation{Institute of Astronomy \& Astrophysics, Nationals
Observatory of Athens, Thessio 11810, Athens, Greece, and
\\Instituto Nacional de Astrof\'isica, \'Optica y Electr\'onica, 72000 Puebla, Mexico}

\author{A. Pouri}
\affiliation{Faculty of Physics, Department of Astrophysics - Astronomy -
Mechanics University of Athens, Panepistemiopolis, Athens 157 83, and 
\\Academy of Athens, Research Center for Astronomy and
Applied Mathematics, Soranou Efesiou 4, 11527, Athens, Greece}

\begin{abstract}
We derive an exact analytical solution for the redshift evolution of
linear and scale-independent bias, by solving a second order differential equation
based on linear perturbation theory. 
This bias evolution model is applicable to all
different types of dark energy and modified gravity models.
We propose that the combination of the current 
bias evolution model with data
on the bias of extragalactic mass tracers
could provide an efficient 
way to discriminate between ``geometrical'' dark energy models and dark energy 
models that adhere to general relativity. 

\end{abstract}
\maketitle

\section{Introduction}
It is well known that the
large-scale clustering pattern of different extragalactic mass tracers
(galaxies, clusters, etc) trace the underlying dark matter distribution
in a biased manner \cite{Kai84} \cite{Bar86}.
Such a biasing is assumed to be statistical 
in nature; with galaxies and clusters being identified as high peaks
of an underlying, initially Gaussian, random density field.
The linear and scale-independent
bias factor, $b$, is thus defined as the ratio of the mass tracer
overdensity to that of the underlying mass  overdensity, or equivalently as the
ratio of the square root of the mass tracer 2-point correlation
function to that of the underlying mass correlation function.
Furthermore, the redshift evolution of bias, $b(z)$, 
is very important in order to relate observations with models of
structure formation and has been shown to be
a monotonically increasing function of redshift.

There are two basic families of analytic bias evolution models. The
first, called the  {\em galaxy merging} bias model, utilizes the halo mass
function and is based on the Press-Schechter \cite{press} formalism, 
the peak-background split \cite{Bar86} and
the spherical collapse model \cite{Cole}.
Many studies have compared the prediction of the {\em merging}
bias model with numerical simulations 
and beyond an overall good agreement, differences have been found in the
details of the halo bias. These differences have lead to modifications
of the original model to include the effects of ellipsoidal collapse
\cite{she01} and to either provide new fitting bias
model parameters \cite{Jing98}, or new forms
of the bias model fitting function \cite{Sel04}
or even a non-Markovian
extension of the excursion set theory \cite{Ma11}.

The second family of bias evolution models assumes a continuous
mass-tracer fluctuation field, proportional to that of the underlying
mass, and the tracers act as ``test particles''. In this context, the
hydrodynamic equations of motion and linear perturbation theory are 
used. This family of models
can be divided into two sub-families: 

\noindent
(a) The so-called {\em galaxy
  conserving} bias model uses the continuity equation and
the assumption that tracers and underline mass share the same velocity
field \cite{Nus94},\cite{Fry96},\cite{Teg98},\cite{Hui07}.
Then the bias evolution is given as the solution of a 1st order
differential equation, and
Tegmark \& Peebles \cite{Teg98} derived: $b(z)=1+(b_{0}-1)/D(z)$, 
with $b_{0}$ is the bias factor at the present time and $D(z)$ the
growing mode of density perturbations.
However, this bias model suffers from 
two fundamental problems: {\it the unbiased problem} ie., the fact that 
an unbiased set of
tracers at the current epoch remains always unbiased in the past, 
and {\it the low redshift problem} ie., the fact that this
model represents correctly the bias evolution only at relatively low
redshifts $z\mincir 0.5$ \cite{Bagla98}. Note that \cite{Simon05}
has
extended this model to also include an evolving mass tracer population
in a $\Lambda$CDM cosmology.

\noindent
(b)
A model based on the basic equation for the evolution of linear density
perturbations, 
and on the assumption of linear and scale-independent bias, 
which are used
to derive a second order differential equation for the bias,
the approximate solution of which provides the evolution of bias 
(see \cite{Bas01} and \cite{Bas08}).
The provided solution applies to
cosmological models, within the framework of general relativity, with a constant in time
dark energy equation of state parameter (ie., quintessence or
phantom). 

In this article, we extend the original Basilakos \& Plionis \cite{Bas01}
bias evolution model to provide an exact solution
valid for all dark energy and modified gravity cosmologies.
This implies that the current bias evolution model 
can be used to put constraints on dark energy 
models as well as to investigate possible departures from 
general relativity. 



\section{The Evolution of the linear growth factor}
In this section, we discuss
the basic equation which governs the behavior of the matter
perturbations on sub-horizon scales and within the framework of any 
dark energy (hereafter DE) model, 
including those of modified gravity (``geometrical dark energy''). 
For these cases, a full analytical 
description can be introduced by considering an extended Poisson equation
together with the Euler and continuity equations.
Consequently, the evolution equation
of the matter fluctuations, for models where the DE
fluid has a vanishing anisotropic stress and the matter fluid is not
coupled to other matter species 
(see \cite{Lue04},\cite{Linder05},\cite{Stab06},\cite{Uzan07},\cite{Linder2007},\cite{Tsu08},\cite{Gann09}), is given by:
\be
\label{odedelta} 
\ddot{\delta}_{\rm m}+ 2H\dot{\delta}_{\rm m}-4 \pi G_{\rm eff} \rho_{\rm m} \delta_{\rm m} =0 
\ee
where $\rho_{\rm m}$ is the matter density
and $G_{\rm eff}(t)=G_{N} Y(t)$, with $G_{N}$ denoting Newton's
gravitational constant.

For those cosmological models which adhere to general relativity,
[$Y(t)=1$, $G_{\rm eff}=G_{N}$], the
above equation reduces to the usual time evolution
equation for the mass density contrast \cite{Peeb93}, while in the
case of modified gravity models (see \cite{Lue04},\cite{Linder2007},
\cite{Tsu08},\cite{Gann09}),
we have $G_{\rm eff}\ne G_{N}$ (or $Y(t) \ne 1$). In this context,
$\delta_{\rm m}(t) \propto D(t)$, where $D(t)$ is the linear growing mode
(usually scaled to unity at the present time). 
Changing variables from $t$ to $a,$ equation (\ref{odedelta}) becomes:
\be
\label{odedelta1} 
\frac{d^{2}\delta_{\rm m}}{da^{2}}+A(a)\frac{d\delta_{\rm m}}{da}-B(a)\delta_{\rm m}=0 
\ee 
where
\be
\label{afun} 
A(a)=\frac{d{\rm ln}E}{da}+\frac{3}{a} \;\;\;{\rm and}\;\;\;
B(a)=\frac{3\Omega_{\rm m}}{2a^{5}E^{2}(a)}\;Y(a)
\ee
with $\Omega_{\rm m}$ being the 
density parameter at the present time and $E(a)=H(a)/H_{0}$ is the
normalized Hubble function.

Useful expressions of the growth
factor have been given by \cite{Peeb93} for the
$\Lambda$CDM cosmology. Several works have also derived the growth factor for
$w(z)=$const DE models (see \cite{Silv94},\cite{Wang98},\cite{Bas03}), and 
for the braneworld cosmology \cite{Lue04}. Also Linder \& Cahn \cite{Linder2007}
derived similar expressions for ``geometrical'' dark
energy models in which the Ricci scalar varies with time,
as well as for models with a time-varying equation of state, while for the
scalar tensor and $f(R)$ models the growth factors are
provided by Gannouji et al. \cite{Gann09} and Tsujikawa et al. \cite{Tsu08}.

\section{The general evolution of bias}
In Basilakos \& Plionis \cite{Bas01}, we assumed 
that for the evolution of the linear bias,
the effects of non-linear gravity and 
hydrodynamics (merging, feedback mechanisms etc) can be ignored 
(see \cite{Fry96},\cite{Teg98}). Then, using linear perturbation theory in
the context of general relativity  [$Y(t)=1$, $G_{\rm eff}=G_{N}$]
we obtained a second order differential equation which describes the 
evolution of the linear bias factor, $b$, between the background
matter and the mass-tracer fluctuation field:
\be\label{eq:hdif} 
\ddot{y}\delta_{\rm m} + 2(\dot{\delta_{\rm m}} + H \delta_{\rm m}) \dot{y} + 4 \pi G_{\rm eff}
\rho_{m} \delta_{\rm m} y =0 \;,
\ee
where $y=b-1$. 
Below, we will prove 
that the above expression is valid for any cosmological
model\footnote{The current theoretical approach 
does not treat the possibility 
of having interactions in the dark sector. Also 
discussions beyond the linear biasing regime can be found in 
\cite{Mc09} (and references therein).} 
including those of modified gravity, with $G_{\rm eff}=G_{N}Y(t)$. 
Since we also make the same assumption, as in our original formulation,
that the tracers and the underlying mass distribution
share the same velocity field and thus the same gravity field,  
the above equation is valid also for cosmological models 
with a modified theory of gravity. Using the latter we have  
\be
\dot{\delta}_{\rm m} + \nabla u \simeq 0 \;\; \mbox{\rm and} \;\; \dot{\delta}_{\rm tr} +
\nabla u \simeq 0 \;,
\ee 
from which we obtain 
\be 
\dot{\delta}_{\rm m} - \dot{\delta}_{\rm tr}=0 \;.
\ee 
Now since we assume linear biasing, we have $\delta_{\rm tr}=b\delta_{\rm m}$, 
and using $y=b-1$, we get that $d(y\delta_{\rm m})/dt=0$. 
Differentiating  the latter twice, we then get:
$\ddot{y} \delta_{\rm m} + 2 \dot{y} \dot{\delta}_{\rm m} + y \ddot{\delta}_{\rm m} =0$.
Solving for $y \ddot{\delta}_{\rm m}$, using the fact that $y
\dot{\delta}_{\rm m} = -\dot{y}\delta_{\rm m}$ and eq.(\ref{odedelta}) we finally obtain eq.(\ref{eq:hdif}).

Transforming equation (\ref{eq:hdif}) 
from $t$ to $a$, we simply derive the evolution equation 
of the function $y(a)$ [where $y(a)=b(a)-1$] which has
some similarity with the form of eq.(\ref{odedelta1}) as expected.
Indeed this is 
\be
\label{ydedelta1} 
\frac{d^{2} y}{da^{2}}+\left[A(a)+\frac{2f(a)}{a}\right]\frac{dy}{da}+B(a)y=0 \;,
\ee 
where $f(a)$
is the growth rate of clustering, a 
parametrization of the linear matter perturbations, given by:
\be
\label{fzz221} 
f(a)=\frac{d{\rm ln}\delta_{\rm m}}{d{\rm ln}a}=\frac{d{\rm ln}D}{d{\rm ln}a}=
\Omega_{\rm m}^{\gamma}(a) \;,
\ee
where $\Omega_{\rm m}(a)=\Omega_{\rm m}a^{-3}/E^{2}(a)$ and
$\gamma$ is the growth index, originally introduced by
Wang \& Steinhardt \cite{Wang98}. Integrating eq.(\ref{fzz221}) we obtain the 
growth factor 
for any type of dark energy: 
\begin{equation}
\label{Dz221}
D(a)=a {\rm e}^{\int_{0}^{a} (dx/x) [\Omega_{\rm m}^{\gamma}(x)-1]} \;.
\end{equation}

In Basilakos \& Plionis \cite{Bas01}, we have provided an approximate
solution of eq.(\ref{eq:hdif}), using $f(z)\sim 1$ (which is valid at relatively 
large redshifts), 
but only in the framework of general relativity, ie., $Y(t)=1$, 
which contains a quintessence (or phantom) dark energy. 
Here our aim is to provide a full analytical solution for 
all possible dark energy cosmologies that have appeared 
in the literature, such as a cosmological constant $\Lambda$
(vacuum), time-varying $w(t)$ cosmologies, quintessence,
$k-$essence, quartessence, vector fields, phantom, 
modifications of gravity, Chaplygin gas etc.

Inserting now $y(a)=g(a)/D(a)$ into eq.(\ref{ydedelta1}) 
and using simultaneously 
equation \ref{odedelta1} and the second 
equality of equation \ref{fzz221}, we obtain: 
\be
\label{qdedelta1} 
\frac{d^{2}g}{da^{2}}+A(a)\frac{dg}{da}=0 \;.
\ee 
That is, the general solution of the latter equation is 
\be
\label{qdelta1} 
g(a)=C_{1}+C_{2} \int \frac{da}{a^{3}E(a)}
\ee
where $C_{1}$ and $C_{2}$ are the integration constants. Utilizing now 
$a=(1+z)^{-1}$, $b=y+1=(g/D)+1$, $b_{0}=b(0)$ and eq.(\ref{qdelta1}), 
we finally obtain the functional form which provides the 
evolution of linear bias for all possible types of DE models,
including those of modified gravity, as:
\be
\label{eq:final} 
b(z)=1+\frac{b_{0}-1}{D(z)}+C_{2} \frac{J(z)}{D(z)} 
\ee
where
\be
\label{eq:final} 
J(z)=\int_{0}^{z} \frac{(1+x) dx}{E(x)}\;.
\ee
Since different halo masses result in different values of $b_{0}$, one should expect 
that the constants of integration $C_{1}=b_{0}-1$ and $C_{2}$ should be
functions of the mass of dark matter halos (see \cite{Bas08}), assuming that 
the extragalactic mass tracers 
are hosted by a dark matter halo of a given mass. Note that an
extension of our model for the case of an evolving mass tracer
population (ie., including the effects of halo merging) is provided in
appendix A.

\begin{figure}[ht]
\mbox{\epsfxsize=8.5cm \epsffile{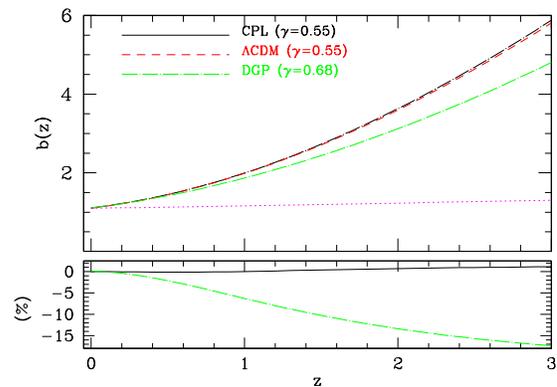}}
\caption{The bias $z$-evolution for different flat 
cosmological models ({\em upper panel}) and their fractional difference with
respect to the $\Lambda$CDM model ({\em lower panel}). The models shown are: 
CPL (solid line) with $w(a)=w_{0}+w_{1}(1-a)$ 
and $\gamma=0.55$,  
concordance $\Lambda$CDM (dashed line) and DGP (dot-dashed) with 
$w(a)=-[1+\Omega_{\rm m}(a)]^{-1}$ and $\gamma=0.68$. Note that we 
use $\Omega_{\rm m}=0.27$, $(w_{0},w_{1})=(-0.93,-0.38)$ \cite{komatsu11}, 
$b_{0}=1.1$ and $C_{2}=0.45$. Finally, we also plot 
(dotted line) the bias evolution for $C_{2}=0$ which corresponds to that
of \cite{Teg98}.}
\end{figure}

Finally, comparing our solution of eq.(\ref{eq:final})
with that of the usual galaxy-conserving bias evolution
model, $b(z)=1+(b_{0}-1)/D(z)$, it becomes evident that the latter
misses one of the two components of the full solution.
Furthermore, our full solution does not suffer from the {\em unbiased} and
the {\em low redshift} problems, but more importantly, the dependence
of our bias evolution model on the 
different cosmologies enters through the
different behavior of $D(a)$, which is affected by $\gamma$ (see equation \ref{Dz221}), 
and of $E(a)=H(a)/H_{0}$. 

It is interesting to mention that measuring
the growth index could provide an efficient 
way to discriminate between modified gravity models and DE models
which adhere to general relativity. Indeed it was theoretically shown that for DE models
inside general relativity the growth index $\gamma$ is well fitted by 
$\gamma_{\rm GR}\approx 6/11$ 
(see \cite{Linder2007},\cite{Nes08}).
Notice, that in the case of the 
braneworld model of Dvali, Gabadadze \& Porrati \cite{DGP} (hereafter DGP)
we have $\gamma \approx 11/16$ (see also \cite{Linder2007}).
Indeed, it has been proposed (see \cite{Vik09}) that an efficient
avenue to constrain the $\gamma$ parameter is by determining
observationally the redshift-dependent linear growth of
perturbations. Alternatively other methods have been proposed in the
literature, such as redshift space
distortions in the galaxy power spectrum and the growth rate of 
massive galaxy clusters (see for example \cite{Guzz08} and references therein).
It is interesting to mention here that the above methods 
also assume a linear and scale-independent bias.

An alternative approach is to use the
current generalized bias evolution,
cosmology and $\gamma$ dependent, relation and
high quality observational bias data to test gravity. 
Of course, the observational
bias data are derived for a particular cosmological model, but it is
an easy task to scale them to each tested model in a consistent
manner. Note, that such data are already 
available in the literature for the case of optical QSOs \cite{Croom}.
If the derived value of $\gamma$ shows scale or time 
dependence or it is inconsistent with $\gamma_{\rm GR}\approx 6/11$,
then this will be a hint that the nature of dark energy reflects 
in the physics of gravity.
Such an analysis is in progress and will be published elsewhere.

In order to visualize the redshift and $\gamma$ dependence of our bias model, we
compare in Fig. 1, a few flat cosmological 
models in which we impose $\Omega_{\rm m}=0.27$, $b_{0}=1.1$ and $C_{2}=0.45$. 
In particular we consider the following cases: 

\noindent
(a) the CPL parametrization \cite{CPL} 
with $\gamma=0.55$ (solid line), 

\noindent
(b) the concordance 
$\Lambda$CDM ($\gamma=0.55$, dashed line), and 

\noindent
(c) the DGP with $\gamma=0.68$ (dot-dashed line). 

The dotted line shows the bias evolution of 
Tegmark \& Peebles \cite{Teg98} model, which 
is also described by our bias model in the limit of $C_{2}=0$.
In the lower panel of Fig.1 we show the fractional difference of the 
model bias with respect to that of the $\Lambda$CDM.

\section{Conclusions}
In this work we provide a general bias 
evolution model, based on linear perturbation theory,
which is valid for  all possible non-interacting dark energy models, including 
those of modified gravity. 
Thus the current generalization of the bias evolution 
model can be viewed as a necessary
step and an ideal tool 
to test the validity of general relativity on cosmological scales.

It is however important to spell out
clearly which are the basic 
assumptions of our model, which are common also to many bias
models in the literature:
(a) the mass tracers and the underline the mass share the same
velocity/gravity field, 
(b) the biasing is linear on the scales of interest (which does not
preclude being scale dependent on small non-linear scales), and 
(c) that each dark matter halo is populated by one
extragalactic mass tracer, which is an assumption that enters, 
at the present development of our model, 
only in the comparison of our model with 
observational bias data and not in the derivation of its functional form.

\vspace {0.4cm}

{\bf Acknowledgments.}  
{\it We thank Joe Silk for useful comments. Manolis Plionis 
acknowledges funding by
Mexican CONACyT grant 2005-49878.}

\appendix
\section{BIAS EVOLUTION FOR AN EVOLVING MASS-TRACER POPULATION}
Here we obtain the 
general linear bias evolution model assuming that the
mass-tracer population evolves with time
according to a $(1+z)^{\nu}$ law.
We now drop the assumption used in section 3, 
that the mass-tracer number density is conserved in time,
by allowing 
a contribution from the corresponding 
interactions among the mass tracers. 
We obtain again the corresponding equation (\ref{eq:hdif}), starting from the continuity
equation and introducing an additional time-dependent term, $\Psi(t)$, which we
associate with the effects of interactions and merging of the mass tracers.
We also make the same assumption, as in our original formulation,
that the tracers and the underlying mass distribution
share the same velocity field (or gravity field). Then:
\be
\dot{\delta}_{\rm m} + \nabla u \simeq 0 \;\; \mbox{\rm and} \;\; \dot{\delta}_{\rm tr} +
\nabla u +\Psi(t) \simeq 0 \;,
\ee 
from which we obtain: 
\be 
\dot{\delta}_{\rm m} - \dot{\delta}_{\rm tr}=\Psi \;.
\ee 
Although we do not have a fundamental theory to model the 
time-dependent $\Psi(t)$ function, 
it appears to depend on the tracer number 
density and its logarithmic derivative as well as on the tracer
overdensity:
$\Psi(t) \propto \Psi(\bar{n}, (1+\delta_{\rm tr}) d \ln {\bar n}/dt)$ (see eq.10 of 
\cite{Simon05} and appendix of Basilakos et al. \cite{Bas08}).

Now, in the context of linear biasing, we have $\delta_{\rm tr}=b \delta_{\rm m}$ 
and utilizing $b=y+1$, we find
that $d(y \delta_{\rm m})/dt=-\Psi$. Differentiating twice the latter we then get:
$\ddot{y} \delta_{\rm m} + 2 \dot{y} \dot{\delta}_{\rm m} + y \ddot{\delta}_{\rm m} =-\dot{\Psi}$.
Solving for $y \ddot{\delta}_{\rm m}$, using the fact that $y
\dot{\delta}_{\rm m} = -\dot{y}\delta_{\rm m}-\Psi$ and equation 
(\ref{odedelta}) we arrive at the following expression:
\be
\ddot{y}\delta_{\rm m} + 2(\dot{\delta}_{\rm m} + H \delta_{\rm m}) \dot{y} + 4 \pi G_{\rm eff}
\rho_{\rm m} \delta_{\rm m} y =-2H\Psi-{\dot \Psi}
\ee
which is the corresponding equation (\ref{eq:hdif}) for the case 
of interactions among the tracers.

Transforming again equation (\ref{odedelta}) 
from $t$ to $a$, we get 
\be
\label{ydedelta11} 
\frac{d^{2} y}{da^{2}}+\left[A(a)+\frac{2f(a)}{a}\right]
\frac{dy}{da}+B(a)y=F(a) 
\ee
where
\be
F(a)=-\frac{2\Psi(a)+a(d\Psi/da)}{a^{2}D(a)H(a)} \;\;.
\ee 
Now, following the same notations ($y=g/D$) as in 
section 3 the above differential equation becomes:
\be
\label{qdedelta2} 
\frac{d^{2}g}{da^{2}}+A(a)\frac{dg}{da}=F(a)  \;. 
\ee 
Integrating eq.(\ref{qdedelta2})
it is straightforward to estimate the general solution
of the bias factor. This is 
\be
\label{qdelta2} 
g(a)=C_{1}+C_{2} \int \frac{da}{a^{3}E(a)}+\int \frac{da}{a^{3}E(a)}
\int^{\alpha} F(\tilde{a})\tilde{a}^{3}E(\tilde{a})d\tilde{a} \;.
\ee
Using the same 
conditions with those provided in section 3, the bias evolution
in the redshift space takes the form
\be
\label{eq:final1} 
b(z)=1+\frac{b_{0}-1}{D(z)}+C_{2}
\frac{J(z)}{D(z)}+\frac{y_{p}(z)}{D(z)}
\ee
where
\be
y_{p}(z)=\int_{0}^{z} \frac{(1+x)}{E(x)} dx
\int_{0}^{x} \frac{F(u)E(u)}{(1+u)^{5}} du\; .
\ee
Obviously, if the interaction among the 
tracers is negligible ($\Psi\simeq 0$) then  
eq.(\ref{eq:final1}) boils down to 
eq.(\ref{eq:final}) as it should.

Now, knowledge of the exact functional form of the interaction
term $\Psi(z)$ would provide the precise redshift evolution of the bias.
As we have analytically proved in the appendix of 
Basilakos et al. \cite{Bas08},  
a reasonable approach regarding the evolutionary $\Psi(z)$ term 
is that: $\Psi(z)=AH_{0}(1+z)^{\nu}$, 
where $\nu \sim 3$. Note that the Hubble constant 
has been maintained for mathematical convenience. 
Inserting the latter equation, $a=(1+z)^{-1}$ 
and $d\Psi/da=-(1+z)^{2} d\Psi/dz$
into the second term of  
eq.(\ref{ydedelta11}) we derive that
\be
F(z)=A(\nu-2)\frac{(1+z)^{\nu+2}}{D(z)E(z)}\;,
\ee
where $A$ is a positive parameter (to be determined
from observational data see Basilakos et al. {\it in preparation}). 
Obviously, for $\nu>2$ the derived bias evolution becomes stronger than 
in the case of no interactions, especially at high redshifts,
which means that due to the merging processes the halos (of some particular 
mass) correspond to higher peaks of the underlying density field with respect 
to equal mass halos in the non-interacting case. 
On the other hand, the $\nu< 2$ case corresponds to the destruction of halos
of a particular mass, which results into a lower-rate of bias evolution with 
respect to the non-interacting case.
Now, for the limiting case with $\nu=2$ we obtain $y_p=0$, 
implying no contribution of the interacting term to the bias evolution 
solution, as in the case with $\Psi=0$, which can be interpreted as 
the case where the destruction and creation processes are counter-balanced.

\end{document}